\documentclass[longbibliography,nofootinbib]{revtex4-1}

\newcommand{\kms}{NuCypher KMS}

\usepackage{listings}
\usepackage{graphicx}
\usepackage{amsmath}
\usepackage[margin=5pt]{subfig}
\usepackage[usenames]{color}

\definecolor{darkgreen}{rgb}{0.00,0.50,0.25}
\definecolor{darkblue}{rgb}{0.00,0.00,0.67}
\newcommand{\figref}[1]{Fig.~\ref{#1}}
\usepackage[breaklinks,pdftitle={NuCypher KMS: decentralized key management system}, pdfauthor={Michael Egorov},colorlinks,urlcolor=blue,citecolor=darkgreen,linkcolor=darkblue]{hyperref}
\usepackage[usenames]{color}
\graphicspath{{pdf/}}

\usepackage[T1]{fontenc}
\usepackage{lmodern}
\begin{document}

\title{\kms: Decentralized key management system}

\author{Michael Egorov}
\email{michael@nucypher.com}
\author{MacLane Wilkison}
\email{maclane@nucypher.com}
\affiliation{NuCypher}

\author{David Nu{\~n}ez}
\email{dnunez@lcc.uma.es}
\affiliation{NICS Lab, Universidad de M{\'a}laga, Spain}

\begin{abstract}
    \kms~is a decentralized Key Management System (KMS) that addresses the limitations of using consensus
    networks to securely store and manipulate private, encrypted data~\cite{cryptoeprint:2017:201}.
    It provides encryption and cryptographic access control, performed by a decentralized network,
    leveraging proxy re-encryption~\cite{wiki:pre}.
    Unlike centralized KMS as a service solutions, it doesn't require trusting a service provider.
    \kms~enables sharing of sensitive data for both decentralized and centralized applications,
    providing security infrastructure for applications from healthcare to identity management to decentralized content marketplaces.
    \kms~will be an essential part of decentralized applications,
    just as SSL/TLS is an essential part of every secure web application.
\end{abstract}

\date{\today}
\maketitle

\tableofcontents

\newpage

\section{Introduction}

\kms~is a decentralized key management system (KMS), encryption, and access control service.
It enables private data sharing between arbitrary numbers of participants in public consensus networks,
using proxy re-encryption to delegate decryption rights in a way that cannot be achieved by traditional symmetric or public-key encryption schemes.
Native tokens are used to incentivize network participants to perform key management and access delegation/revocation operations.

\subsection{Background}
\begin{figure}
    \centering
    \subfloat[Centralized KMS]{\includegraphics[width=0.4\columnwidth]{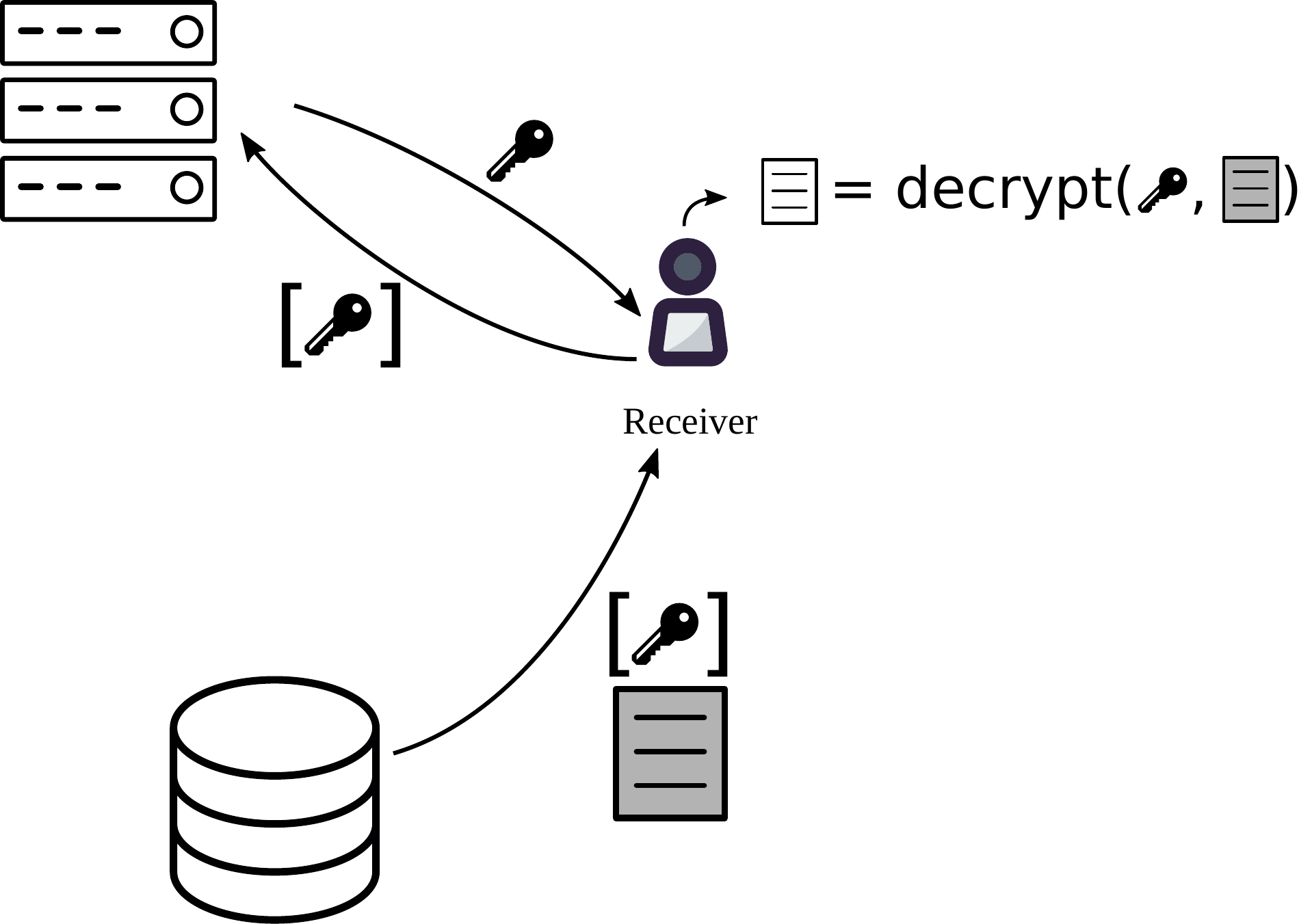}}
    \qquad
    \qquad
    \subfloat[PRE-based KMS]{\includegraphics[width=0.4\columnwidth]{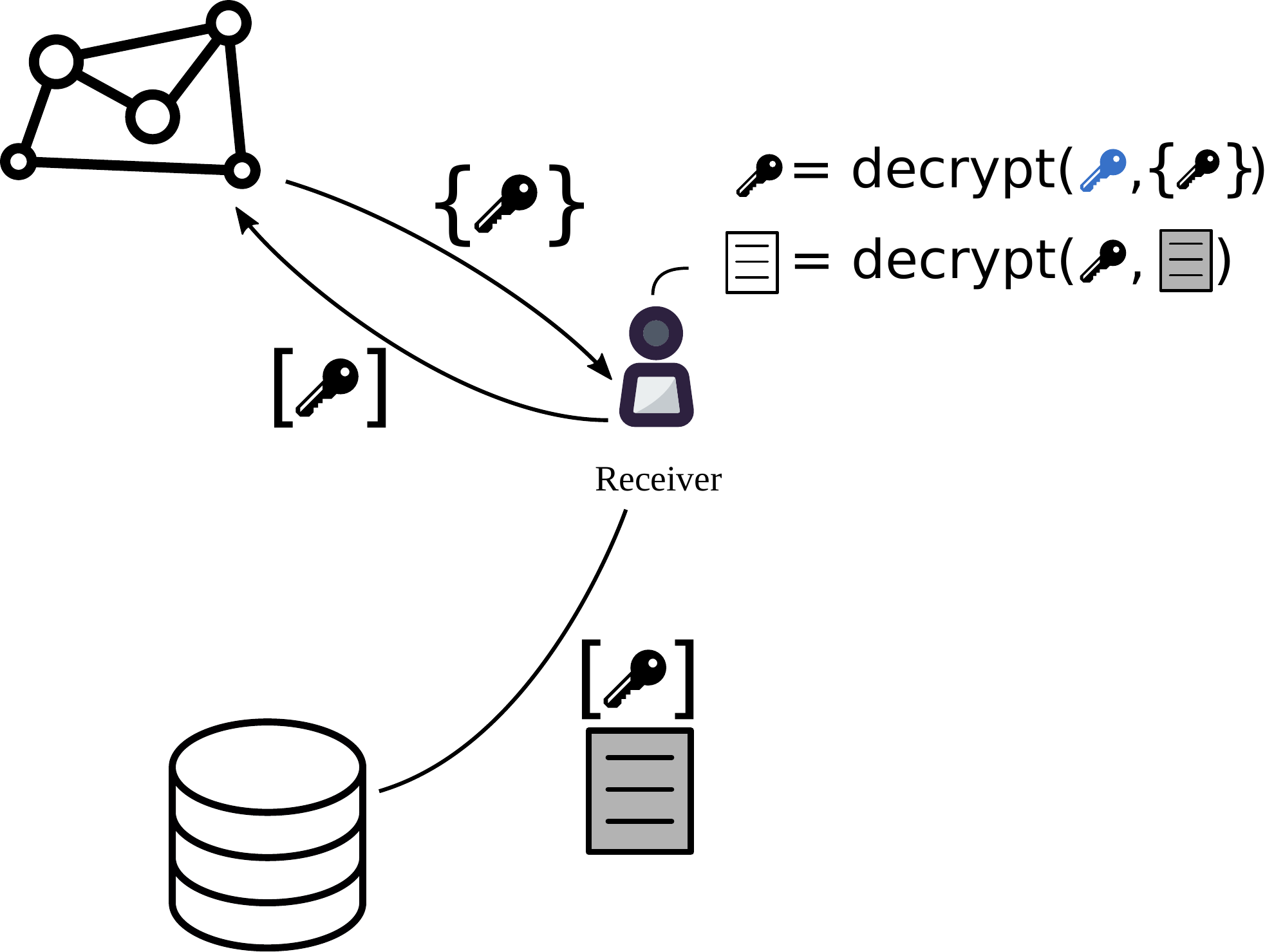}}
    \caption{Difference between a centralized key management system (KMS) and the one which uses proxy re-encryption (PRE)}
    \label{fig:kms}
\end{figure}

A key management system (KMS) is an integrated approach for generating, distributing, and managing cryptographic keys for devices and
applications~(\figref{fig:kms}).
A KMS includes the backend functionality for key generation, distribution, and rotation as well as the client functionality for
injecting, storing, and managing keys on devices~\cite{wiki:kms}.

As the root of trust, it's critical that a KMS is appropriately configured, managed, and protected.
Historically, this has meant deploying a KMS on-premises in hardware security modules (HSM)~\cite{wiki:hsm} or using tools like
HashiCorp's Vault~\cite{web:hashicorp-vault}.
However, this requires a high degree of technical sophistication as well as upfront capital investment.
To ease the technical burden and provide more competitive pricing, vendors like Amazon CloudHSM~\cite{web:aws-cloudhsm},
Google Cloud KMS~\cite{web:google-cloud-kms}, Azure Key Vault~\cite{web:azure-key-vault} and TrueVault~\cite{web:truevault}
have begun offering KMS as a service.
However, KMS as a service offerings necessitate placing an undue level of trust in the service provider, which may
be inappropriate for security-critical applications.

Public consensus networks, such as Bitcoin and Ethereum, are a promising solution to this centralization problem.
But the limitations of public consensus networks in performing cryptographic operations that involve the manipulation of secret
data are well-established~\cite{cryptoeprint:2017:201}. Consensus networks employ a volunteer network of nodes,
which is subject to constant churn and not as reliable as central infrastructure when it comes to availability and
enforcing access management rules.

\kms~ uses a decentralized network to remove the reliance on central service providers, proxy re-encryption for cryptographic
access control, and a token incentive mechanism to ensure reliability, availability, and correctness.
Because of the use of proxy re-encryption, an unencrypted symmetric key (which gives the ability to decrypt private data) is never exposed
server-side~(\figref{fig:kms}), and there is no single point of security failure.
Even if compromised, hackers would only get re-encryption keys but access to the file is still protected.

\section{Architecture}

\subsection{Cryptographic primitives}

\subsubsection{Symmetric encryption}

Symmetric or secret key encryption requires users to know a common secret key.
For convenience, we refer to this common secret key as DEK (data encryption key).

Two operations are defined for symmetric ciphers:
\begin{align}
    c &= \text{encrypt}_{sym}(dek, d);\\
    d &= \text{decrypt}_{sym}(dek, c);
\end{align}
where $d$ is plaintext data, and $c$ is ciphertext (encrypted data).

The most useful symmetric key encryption algorithms for our purposes are AES (because it's normally hardware-accelerated)~\cite{wiki:aes}
and Salsa20~\cite{wiki:salsa20}.

Symmetric block ciphers can operate in different modes.
We use modes of operation that yield probabilistic encryption (such as GCM for AES), to guarantee strong semantic security.
For simplicity, we omit details about the particular modes of operations, treating the value of \emph{nonce} related to the mode of operation as part of the
ciphertext $c$.

\subsubsection{Public-key encryption}

Public-key encryption (PKE) is a type of encryption where two parties (a sender and a receiver) exchange information without any required common secret.
Every participant has a key pair (a public key $pk$ and a secret/private key $sk$).
If the sender has a key pair $sk_s/pk_s$ and the receiver has a key pair $sk_r/pk_r$, the sender can encrypt a message with the receiver's public key,
and the receiver can decrypt with his secret key.

Hybrid cryptosystems can be created that combine the efficiency of symmetric encryption with the convenience of PKE.
A hybrid cryptosytem encryption flow defines the functions:
\begin{align}
    dek &= \text{random}();\\
    c &= \text{encrypt}_{sym}(dek, d);\\
    edek &= \text{encrypt}_{pke}(dek, pk_r).
\end{align}
The pair $(edek, c)$ is used to transform the encrypted data.
The associated decryption by the receiver defines the functions:
\begin{align}
    dek &= \text{decrypt}_{pke}(edek, sk_r);\\
    d &= \text{decrypt}_{sym}(dek, c).
\end{align}

\subsubsection{Proxy re-encryption}
\begin{figure}
\centering
    \includegraphics[width=0.6\columnwidth]{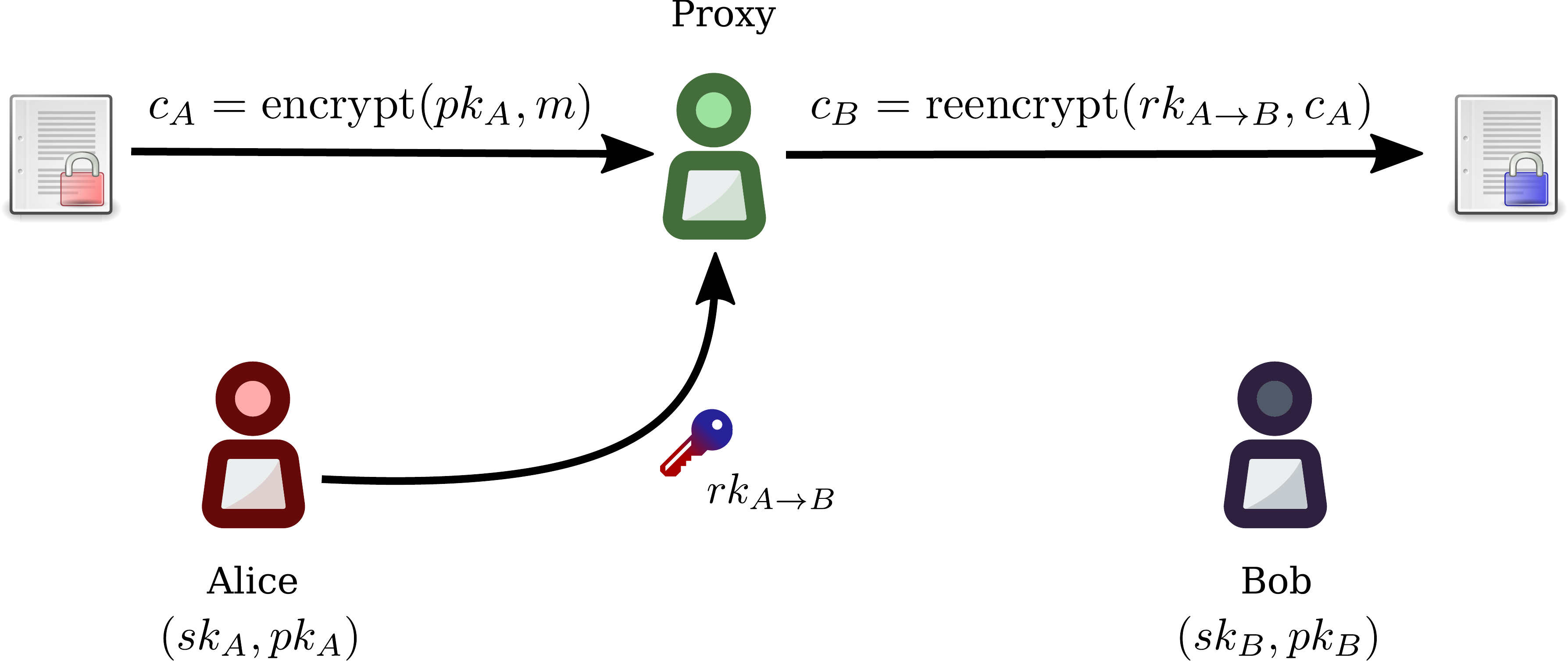}
    \caption{Main actors and interactions in a PRE environment}
    \label{fig:pre}
\end{figure}

Proxy re-encryption (PRE)~\cite{wiki:pre,nunez2017proxy} is a type of public-key encryption (PKE) that allows a proxy entity to transform ciphertexts
from one public key to another, without learning anything about the underlying message~(\figref{fig:pre}).

In a typical proxy re-encryption scenario, there are several actors. 
Alice, the data owner, has public key $pk_A$. Any entity that knows this key can produce encrypted data that only she can decrypt, using her secret key $sk_A$. 
Suppose that a data producer encrypts a message $m$, with Alice's public key $pk_A$, 
resulting in ciphertext $c_A$.
Alice decides to delegate access to message $m$ to Bob, who has the key pair $(pk_B, sk_B)$.
To do so, Alice creates a re-encryption key:
\begin{equation}
    rk_{A\rightarrow B} = \text{rekey}(sk_A, pk_B).
\end{equation}
Importantly, in single-use, uni-directional proxy re-encryption schemes, this re-encryption function is one way, and $rk_{A\rightarrow B}$ cannot be decomposed into its component parts
(at least, without also knowing $sk_A$ or $sk_B$).
All it can do is re-encrypt $c_A$ such that it is transformed into $c_B$:
\begin{equation}
    c_B = \text{reencrypt}(rk_{A\rightarrow B}, c_{A}).
\end{equation}
Bob can then decrypt $c_{B}$ using his secret key $sk_{B}$.

Compared to existing PKE protocols which are ideal for 1-to-1 communication, PRE is more scalable for N-to-N communication
with arbitrary numbers of data producers and consumers.
It doesn't require knowing the recipient of a message in advance, as the re-encryption token can be created and applied at any point.
This makes it well-suited for distributed systems such as blockchain, IoT, and big data~\cite{web:nucypher-hadoop}.

There are many proxy re-encryption algorithms with different properties.
For the first version of \kms~we choose ECIES for which we created a proxy re-encryption algorithm~\cite{umbral-spec}.
It is very similar to the most simple (and performant) algorithm,~--- BBS98~\cite{BBS98}, but provides better security guarantees.
However, sometimes we want to delegate re-encryption to multiple nodes, in order to split the trust between them and apply time-based or conditional
re-encryption policies.
In this case we use the AFGH scheme~\cite{AFGH}.
Quantum-resistant NTRU may also be used if the need arises~\cite{wiki:ntru,ntrureencrypt}.

Just like BBS98, our proxy re-encryption for ECIES creates a re-encryption key out of two secret keys, rather than a secret+public key.
However, we don't want the sender to know the receiver's secret key.
To get around this problem, we randomly generate an ephemeral key pair $sk_e/pk_e$.
Then, access delegation looks like:
\begin{align}
    \label{eq:ephemeral-trick}
    sk_e &= \text{random}();\\
    rk_{A\rightarrow e} &= \text{rekey}(sk_A, sk_e);\\
    \label{eq:ephemeral-trick-end}
    sk_e^{\prime} &= \text{encrypt}_{pke}(pk_B, sk_e);\\
    rk_{A\rightarrow B} &= (rk_{A\rightarrow e}, sk_e^{\prime});
\end{align}

The re-encryption node uses $rk_{A\rightarrow e}$ to re-encrypt any ciphertext $c_A$ (whose underlying message is $m$) so it can be decrypted by $sk_e$. Since the receiver needs $sk_e^{\prime}$ to decrypt the re-encrypted ciphertext, this is attached to the re-encryption result. Therefore, the re-encryption process is as follows:

\begin{align}
    c_e = \text{reencrypt}(rk_{A\rightarrow e}, c_{A});\\
    c_B = (c_e, sk_e^{\prime});\\
\end{align}


The decryption by the receiver will then look like:
\begin{align}
    sk_e &= \text{decrypt}_{pke}(sk_B, sk_e^{\prime});\\
    m &= \text{decrypt}_{pke}(sk_e, c_e);\\
\end{align}

This approach was mentioned beforehand by Ateniese et al.~\cite{AFGH}, and also beyond proxy re-encryption work.
The scheme is not considered to be ``key-optimal'', however, in our case, has a very competitive performance.

\subsection{A brief review of re-encryption schemes}

There are different properties which re-encryption schemes may enjoy.
In this subsection, we consider several of them relevant to our use cases.
A full survey with consideration of all the known proxy re-encryption algorithms and their properties was published recently~\cite{nunez2017proxy}.

One of the important properties to consider is whether an algorithm is interactive or not.
``Interactive'' means that a re-encryption key is computed out of two secret keys:
\begin{equation}
    re_{ab} = \text{rekey}(sk_a, sk_b).
\end{equation}
``Non-interactive'' means that one needs to know only the owner's private key and the delegatee's public key:
\begin{equation}
    re_{ab} = \text{rekey}(sk_a, pk_b).
\end{equation}
Examples of interactive algorithms are: BBS98~\cite{BBS98}, our ECIES re-encryption (based on BBS98), and LWE-based re-encryption~\cite{lwe-reencryption}.
An example of a ``non-interactive'' algorithm is AFGH~\cite{AFGH}.
Even though it initially seems that we would only be interested in non-interactive algorithms,
we can tweak interactive algorithms on the protocol level to share with a public rather than a private key by 
using ephemeral keys (Eqs.~\ref{eq:ephemeral-trick}-\ref{eq:ephemeral-trick-end}).

Another interesting property is whether an algorithm is uni-directional or bi-directional.
Bi-directionality means that it is possible to compute $re_{ba}$ from just $re_{ab}$.
In uni-directional algorithms that is impossible.
Bi-directional algorithms are effectively uni-directional when using ephemeral keys
(Eqs.~\ref{eq:ephemeral-trick}-\ref{eq:ephemeral-trick-end}).
Nevertheless, in Sec.~\ref{sec:hierarchical-data} we show that uni-directionality can be very convenient for making complex hierarchical data shareable in a
scalable way.
An example of a bi-directional algorithm is BBS98~\cite{BBS98}.

Proxy re-encryption algorithms can also be single-hop or multi-hop.
``Multi-hop'' means that if we have $re_{ab}$ and $re_{bc}$, these two re-encryption keys can be applied in series to convert ciphertext $c_a$ to a ciphertext
$c_c$ without the participation of $b$:
\begin{equation}
    c_c = \text{reencrypt}(re_{bc}, \text{reencrypt}(re_{ab}, c_a)).
\end{equation}
Sometimes, it is even possible to compute $re_{ac}$, which for BBS98 is:
\begin{equation}
    re_{ac} = re_{ab} \cdot re_{bc}.
\end{equation}
``Single-hop'' means that if $c_b$ is obtained via re-encryption, it is impossible to re-encrypt it further.
Multi-hop schemes are useful for key rotation (Sec.~\ref{sec:key-rotation}) and for our hierarchical data sharing scheme (Sec.~\ref{sec:hierarchical-data}).
The ephemeral key trick (Eqs.~\ref{eq:ephemeral-trick}-\ref{eq:ephemeral-trick-end}) is, effectively, single-hop
(although key rotation is still possible because ephemeral keys are only created when data is shared with other parties).
Examples of multi-hop algorithms include BBS98~\cite{BBS98} and the LWE-based algorithm~\cite{lwe-reencryption}.
The AFGH algorithm~\cite{AFGH} is single-hop.

Another seemingly important property is collusion resistance.
Informally, collusion resistance means that having just $re_{ab}$ and $sk_b$ it is impossible to derive $sk_a$. 
Conversely, a lack of collusion resistance makes it possible to obtain $sk_a$ from $re_{ab}$ and $sk_b$.
At first glance this seems to be very desirable for security.
However, even with collusion resistant algorithms, if an attacker has both $re_{ab}$ and $sk_b$, he is quite capable of re-encrypting and reading any data that, originally, is decryptable by $sk_a$. 
Thus, collusion resistance only becomes important when the same keypair is used for other purposes (e.g., signing, key derivation, etc). Nonetheless, it is good practice to use separate key pairs for both functions. 
For our purposes, collusion resistance appears not a priority because we don't have another uses for the delegator's keypair apart from encryption. 
Collusion resistant algorithms include AFGH~\cite{AFGH} and LWE-based re-encryption~\cite{lwe-reencryption}.
Non-collusion resistant algorithms are BBS98~\cite{BBS98} and ECIES re-encryption.

We may need to make it impossible to identify producers and/or consumers of the data from re-encryption keys~(Sec.~\ref{sec:anonymity}).
Often, a proxy who knows all public keys in the system is able to deduce this information. 
Many PRE schemes are not anonymous from the re-encryption point of view ~\cite{BBS98,AFGH}. However, the LWE-based re-encryption scheme ~\cite{lwe-reencryption} enjoys the property of anonymity.

Finally, as for any cryptosystem, notions of CPA-security~\cite{wiki:cpa} (security against chosen-plaintext attacks)
and CCA-security~\cite{wiki:cca} (security against chosen-ciphertext attacks) are applicable to proxy re-encryption.
All of the algorithms we have mentioned so far are only CPA-secure, except ECIES re-encryption and LWE, which are also CCA-secure~\cite{lwe-reencryption}.

\subsection{Signing encrypted messages}

In public-key encryption algorithms, anyone can encrypt using $pub_a$.
While this is useful, it also allows for malicious users of the network to encrypt data as if they were $A$.
So the data has to be signed in order to prove the identity of the sender to the recipient.

However, we want to make it possible to anonymize the protocol~(Sec.~\ref{sec:anonymity}) because
a public digital signature that authenticates the owner of the data also raises the possibility of a re-encryption node attempting to extort money from the owner.
As such, it makes sense to:
\begin{itemize}
    \item include a digital signature as part of the plain text, so that it is impossible to verify the signature until the ciphertext is decrypted,
    \item use different key pairs for signing and for encryption (especially considering the lack of collusion resistance of some re-encryption
        cryptosystems).
\end{itemize}

\subsection{Re-encryption nodes}

\begin{figure}
\centering
    \includegraphics[width=0.4\columnwidth]{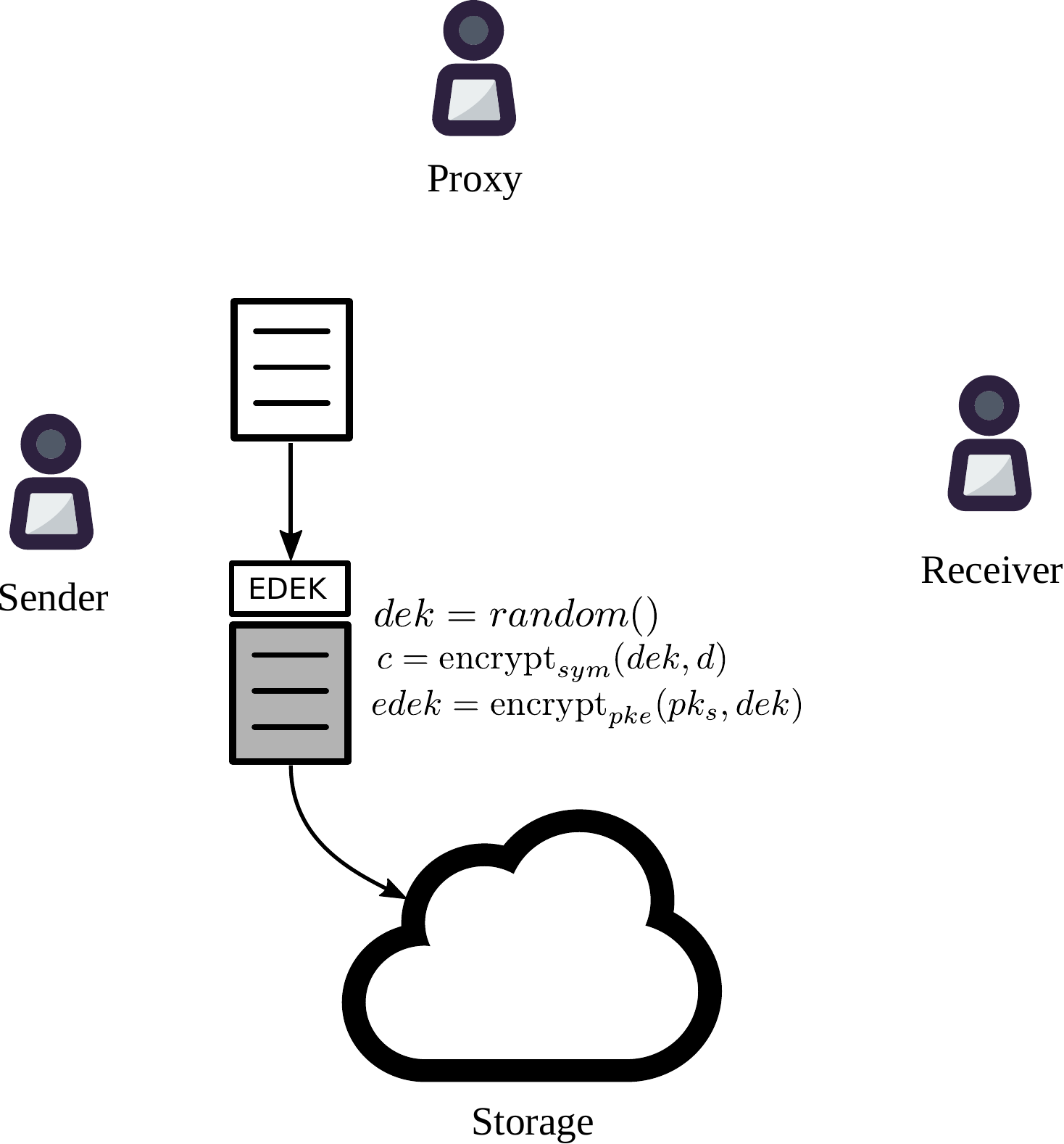}
    \caption{Architecture: encryption}
    \label{fig:arch-encrypt}
\end{figure}

When the data is stored in a cloud or decentralized storage, it is encrypted with the data owner's (\emph{sender}) key $pk_s$~(\figref{fig:arch-encrypt}).
The data itself is encrypted with a random symmetric key $dek$, with one key per file.
The $dek$, encrypted with $pk_s$ is attached to the encrypted data.
This combination $(edek, c)$ can be stored anywhere - in IPFS, Swarm, S3, or any kind of decentralized or centralized storage.

\begin{figure}
\centering
    \includegraphics[width=0.4\columnwidth]{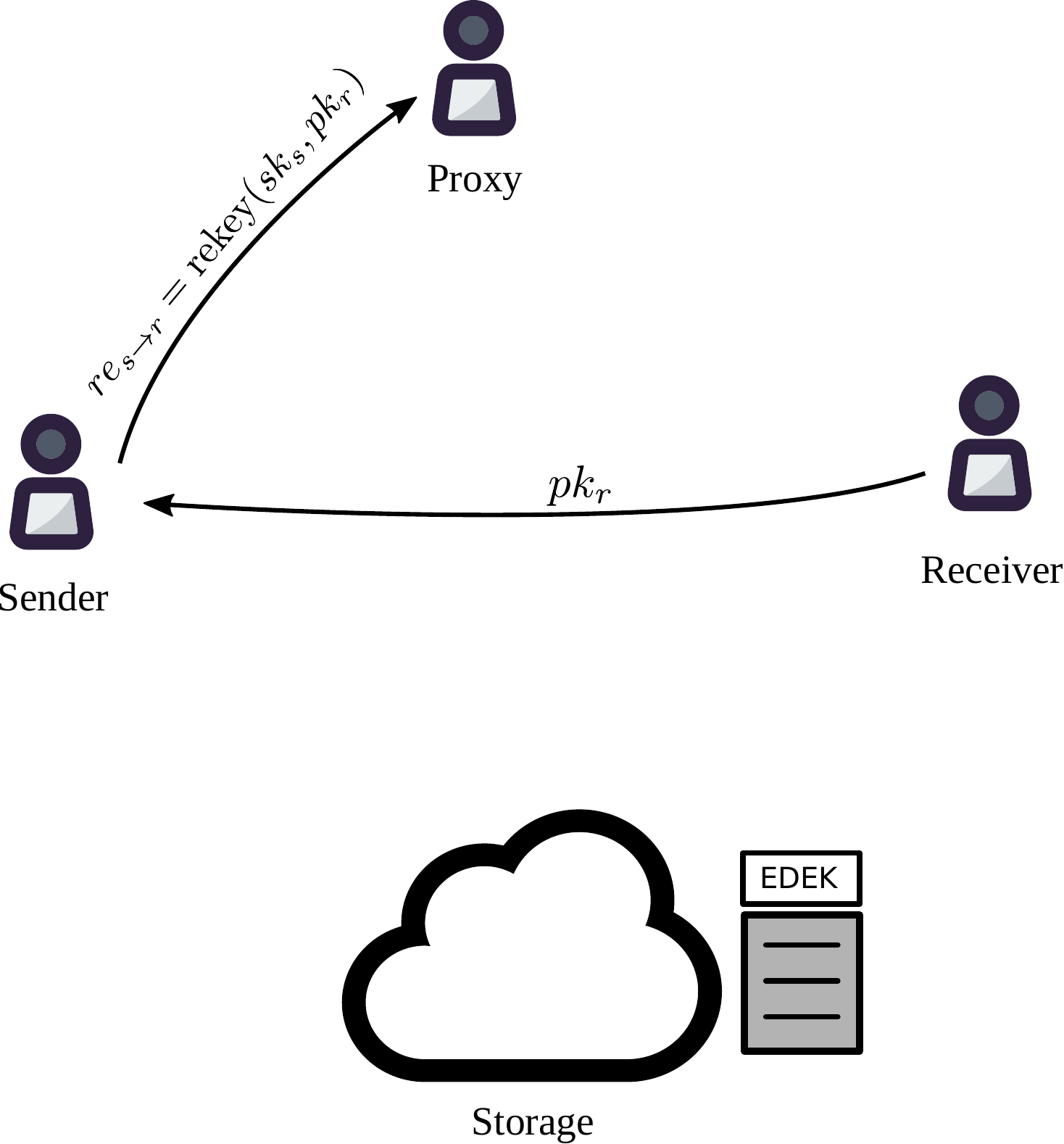}
    \caption{Architecture: access delegation}
    \label{fig:arch-delegate}
\end{figure}

When storing the data, the user to whom we delegate access is not necessarily known in advance.
First, the receiver should show the sender his public key~(\figref{fig:arch-delegate}).
It often makes sense for the public key to correspond to an address in the Ethereum network (to prove a payment has been made from that
address for a digital content subscription, for example).
The sender generates a re-encryption key $re_{s\rightarrow r}$ (including an encrypted random ephemeral key when needed) and sends it to a random re-encryption
node, selected according to proof-of-stake out of the active nodes in a decentralized network.
The case where multiple nodes are selected for redundancy or security will be discussed later.
The nodes which have shared data of user \emph{sender} with a user \emph{receiver} register this information in the network.

\begin{figure}
\centering
    \includegraphics[width=0.6\columnwidth]{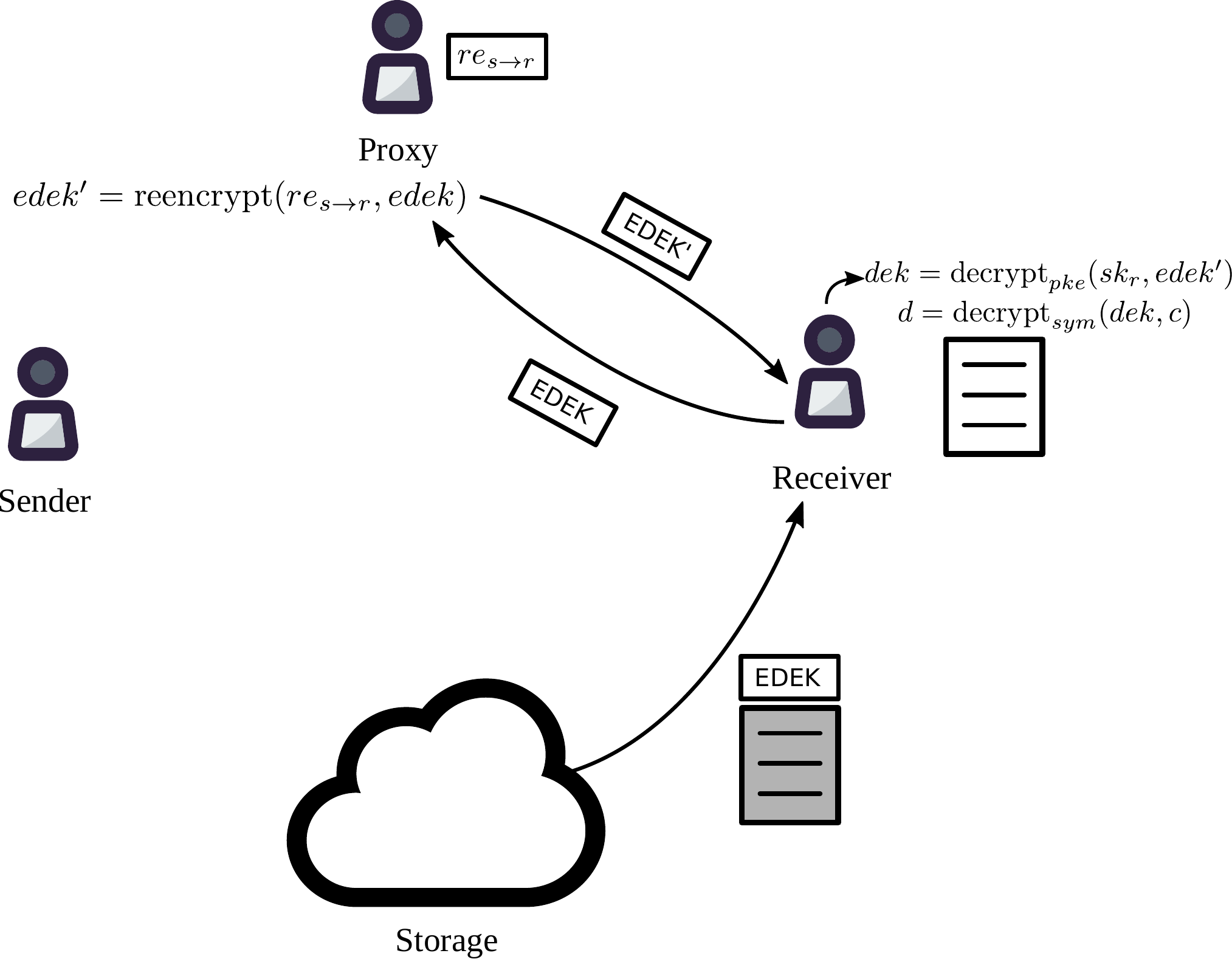}
    \caption{Architecture: decryption}
    \label{fig:arch-decrypt}
\end{figure}

When the receiver wants to decrypt data shared with him, he first downloads that data from storage or an encrypted stream~(\figref{fig:arch-decrypt}).
He separates out $edek$ from the message and sends $edek$ to the network of re-encryption nodes and finds active re-encryption nodes which can share the data
of the sender with the receiver (those which have re-encryption key(s) $re_{s\rightarrow r}$).
The receiver asks the node(s) that has the re-encryption key to transform $edek$ to $edek^{\prime}$ and uses his own secret key $sk_r$ to decrypt it and
obtain DEK.
Now, he can use DEK to decrypt the bulk of the data.

\section{Network security}

In the network, there are multiple re-encryption nodes which apply access management policies.

Proxy re-encryption allows \kms~to split the trust between access management and decryption rights,
without introducing an always-online always-trusted entity (such as a traditional key management system).
Miners never see plaintext data, or anything which allows them to decrypt the data.
They are solely responsible for storing re-encryption keys and applying re-encryption functions.

The first risk of this model is collusion between a miner and a reader of the data.
If the miner gives the reader re-encryption keys for the data,
the data can be decrypted at any time by the data reader, circumventing any conditional or time-based constraints.
We counteract this threat in multiple ways: pseudo-anonymity of re-encryption keys, split-key proxy re-encryption, and a challenge protocol.
In addition, we apply economic incentives for fair operation, described in Sec.~\ref{sec:token}.

The second risk is nodes malfunctioning (returning fake data instead of performing re-encryptions).
We solve this problem using a challenge protocol.

The third risk is nodes colluding with each other to perform 50\% attacks.
This risk is usually deadly for multi-party computations~\cite{vitalik-secret-dao} (such as Enigma~\cite{enigma}). However, in our case the attacker only gains the ability to wrongfully apply re-encryption policies, not to decrypt data nor to grant access to a user who hasn't been granted access to the data.
Ideally, the system should be as decentralized as possible, however 50\% attacks don't compromise the confidentiality of the data, just like 50\% attacks in
proof-of-work cryptocurrencies don't give an attacker the ability to move funds.

\subsection{Pseudo-anonymity}
\label{sec:anonymity}

It is highly beneficial for the security of the system that re-encryption nodes do not know what it is they are re-encrypting.
This prevents them from knowing which re-encryption keys to perform collusion attacks on (and trying to collude with all the network participants
is infeasible when the network is decentralized).
But pseudo-anonymity of re-encryption keys also enables us to run a challenge protocol (Sec.~\ref{sec:challenge-protocol}).

We leave designing an anonymous protocol for re-encryption as future improvements.
However, we point out the following properties it should (and should not) possess.
Firstly, the re-encryption scheme should be key-private~\cite{Ateniese-key-private,lwe-reencryption}.
Otherwise, it would be possible to determine the ownership of the key by iterating over pairs of all the known public keys.
Secondly, the re-encryption node and the recipient of the data should not have the same identifier for the same re-encryption key.
This rules out a simple way of storing a re-encryption key in a key-value store while the recipient can come up with the key.

\subsection{Split-key re-encryption}

Imagine that a re-encryption node decides to re-encrypt data immediately rather than to apply conditional policies as instructed.
A split-key proxy re-encryption scheme can be used to solve this problem.

Instead of one re-encryption key, m-of-m re-encryption keys can be used to produce ``re-encryption shares.''
These shares can be combined client-side.
An m-of-m scheme exists for AFGH~\cite{AFGH} encryption.
A collusion attack here would require $m$ miners \emph{and} the reader of the data.

A threshold-based m-of-n scheme, similar to what will soon be published by NICS Labs from University of M{\'a}laga, appears to be even more appropriate for this task.

\subsection{Challenge protocol}
\label{sec:challenge-protocol}

There is a risk of miners returning random numbers instead of correctly re-encrypting data.
Since the data is private, users of the system cannot publish this data and their key as proof that the miner has cheated.

It is impossible for a miner to distinguish between a ``true re-encryption'' and a re-encryption of random data.
So, we can produce a number of ``fake'' re-encryption keys which are designed specifically to challenge the miners.
If a miner cheats, the data and the key for this challenge aren't associated with any private data.

The miners should show the hashes of data before and after re-encryption to the network.
If this re-encryption was a challenge and the miner has cheated, challengers can present a proof that non-sensitive keys related to this challenge should actually
produce a different re-encryption result, and the miner's collateral deposit can be awarded to the challenger.

The system should also intentionally produce a number of ``wrong re-encryptions'', in order to incentivize challengers to operate, as pointed out by
Truebit~\cite{truebit}.

Designing a challenge protocol is a complex problem related to ``fair exchange'' protocols~\cite{BitcoinMPC2016,Bentov2014,AccountableStorage}.
It requires careful design and testing, and Ethereum's Proof-of-Stake (Casper) protocol is facing this complexity now.
It may be possible to just check correctness on the level of the encryption algorithm~\cite{Zhou2002}.

Special consideration should be given to protecting re-encryption keys from leaking.
The following challenge protocol is proposed.

When accepting responsibility for a re-encryption key, a miner expects to get a fee $f$ over time $T$, so the owner of the data deposits $f$ coins.
The miner should also put up collateral $c$ which will be forfeit if leaks the re-encryption key is leaked.

If a challenger proves that the miner has leaked a re-encryption key, the challenger should be rewarded.
However, the data owner may challenge the miner in order to fraudulently collect the challenge reward.
We make this ``self-challenge'' infeasible.
If the challenge has happened after time $t$, the challenger will get
$\alpha f t/T$ coins, where $\alpha < 1$.
The data owner in this case gets $(1 - t/T) f$ coins returned.
The collateral and the rest of the fee gets seized for the benefit of the other participants of the network, with the total amount of $c + (1 - \alpha) t/T$.

There also should be no incentive for the owner of the data to fake-challenge the miner instead of revoking the policy.
So, in a ``correct'' revocation, the owner of the data gets $(1 - t/T) f$ coins back, and the miner gets $c + ft/T$ coins, where $c$ is the collateral which
was staked.

\subsection{Relevancy of possible threats to different use cases}

In a mobile device management use case~(Sec.~\ref{sec:mdm}),
the most important thing is to revoke access from a lost or stolen device before the data is compromised.
Imagine a possible attack where someone steals the device and colludes with the relevant miners.
As such, there must be no way for miners to identify a user, and vice versa.
Another possible attack is a group of miners revoking access and demanding additional payment to re-encrypt.
However, there is no incentive to do so since the owner of the data can easily re-grant access to that mobile device.
Another possible threat is a mining node keeping re-encryption keys for a long time beyond the life of the policy,
waiting for someone to attack the end-user device and collude with the mining node.
In order to prevent this threat, it is important that the mining node cannot figure out if the data is valuable or not,
and a good way to do this is to anonymize the data owner and the data itself.
In other words, we prevent the mining node from discovering which data the $edek$ corresponds to.

Decentralized DRM~(Sec.~\ref{sec:drm}) assumes that once content (a file or a piece of video) is decrypted, it has been purchased,
so access revokation is not really an issue.
However, if a node knows that the content is very expensive, they may attempt to approach the buyer and solicit a cheaper price for the
content, cutting out the original seller.
To prevent this, we should anonymize the recipient of the data.
It would also help to hide the exact pricing information from the mining node while still allowing it to verify the necessary amount was paid using
zk-SNARKs~\cite{consensys-snarks}.

When \kms~is used to secure access to files~(Sec.~\ref{sec:files}) or messages, both granting and revocation of access are important.
So full anonymization is highly desirable.
Possible attacks include recipients of the data bribing miners to continue having access after it should have been revoked and
miners extorting fees from the owner when it is critical to revoke access.
Anonymization appears to be an important part of making such attacks infeasible.

\subsection{Hardware-enforced security}

If miners misbehave, they risk losing their collateral deposit.
However, rather than purposeful ill-intent, miner nodes could be the victim of a third-party attack.
In order for miners to prevent their nodes from being compromised, they can use trusted computing on commonly accessible secure hardware~\cite{Yang2011},
such as the latest generation of Intel CPUs (Skylake+) which support SGX, or NVidia GPUs.

Intel SGX technology~\cite{wiki:sgx} promises to run \emph{any} computations in a secure environment.
It was previously proposed to have a decentralized network for managing secrets relying on the SGX technology rather than fairness of
miners or re-encryption~\cite{sgx-blockchain-encryption}.

Alternatively, holding and applying re-encryption keys can be done inside GPUs.
It was shown that GPUs can serve as trusted platform modules with limited capabilities (although they can certainly execute
re-encryption)~\cite{gpu-trusted}.

\section{Performance considerations}

\section{Functionality}

\subsection{Local encryption library and daemon}

\kms~can be interfaced from a traditional, centralized application.
In Python, sharing a file in IPFS would look like:
\begin{lstlisting}[frame=single,language=Python]
import nkms
nkms.connect()  # Using default config
path = 'ipfs://QmTkzDw.../to_the_moon.avi'
nkms.share('0xab12...', path, time=86400)
\end{lstlisting}
If there is no library client available, there could be a local API server, similar to how geth is used to interact with the Ethereum network.

Reading this file would be:
\begin{lstlisting}[frame=single,language=Python]
import nkms
client = ipfsapi.Client(...)
nkms.connect()  # Using default config
path = 'ipfs://QmTkzDw.../to_the_moon.avi'
edata = client.cat(path)
data = nkms.decrypt(edata, path)
print(data)
\end{lstlisting}
The ``decrypt'' function splits edata into $edek$ and the actual encrypted data,
requests the kms network to transform $edek$ into $edek^{\prime}$,
decrypts $edek^{\prime}$ with the receiver's private key and
decrypts data with the obtained $dek$.

The preliminary version of the API will include the following functions:
\begin{itemize}
    \item connect~--- connect to the decentralized network,
        taking configuration from arguments or a config file;
    \item write~--- encrypt data with a public key corresponding to the owner of the file and save in a storage backend;
    \item read~--- download data from the storage backend, ask the decentralized network to re-encrypt, and decrypt with our private key;
    \item delete~--- delete file and re-encryption keys associated with it;
    \item decrypt~--- decrypt data which we've already read;
    \item split-edek~--- low level function to separate out encrypted symmetric encryption key from the data;
    \item share/renew/revoke~--- create a permission to read all the data we own, or a subset of it, based on file path or policy.
        The policy can include time limits and other conditions;
    \item read-policies/update-policies/delete-policies~--- read and change all the access policies we've created.
\end{itemize}

\subsection{Sharing short secrets}

Low-level functionality allows encrypting and delegating access to binary secrets (such as database credentials) or groups of these secrets without
storing them in individual files.
A possible backend for storing simple secrets like these could be a hierarchical config file (in YAML or JSON format) stored in IPFS,
or even simply in the same instance image.
Client-side software can parse this file and ask to re-encrypt only the secrets it needs.
There could be per-field and per-subfield re-encryption keys for granular permissions.

\subsection{Sharing files and hierarchical data}
\label{sec:hierarchical-data}

When working with files, each file and each directory is encrypted under its secret key.
Each file has its DEK stored encrypted by its secret key, the directory's secret key, and so on, up to the root folder of the user.
When a directory is shared, a re-encryption key is created only for the shared directories.

\begin{figure}
\centering
    \includegraphics[width=0.45\columnwidth]{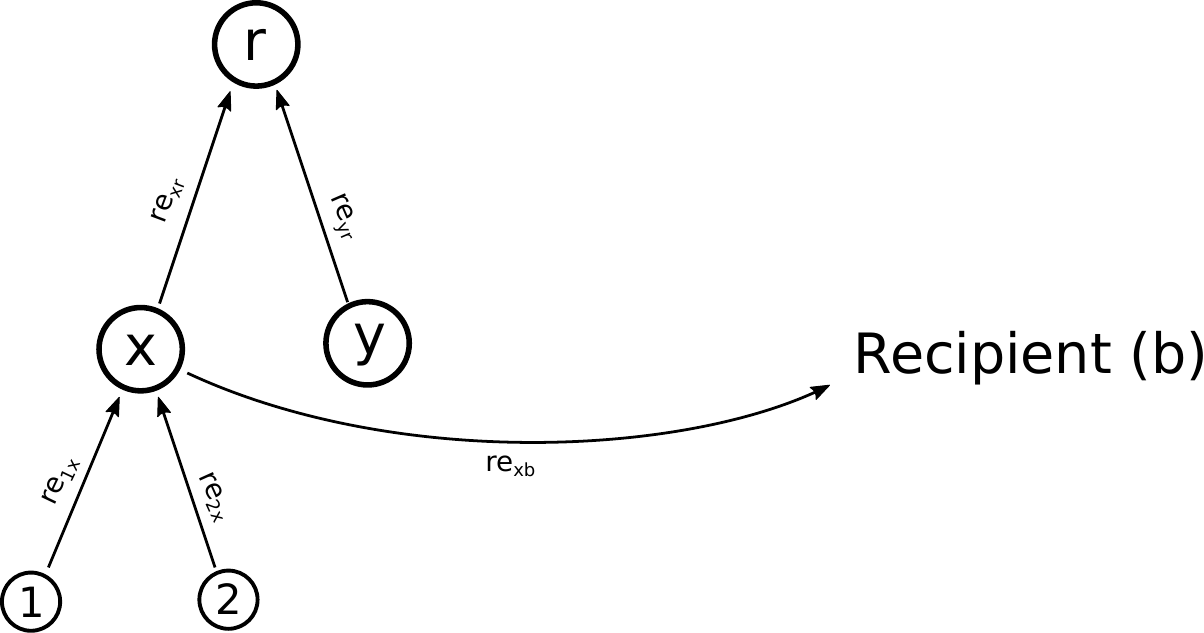}
    \caption{Sharing hierarchical structures (such as files and folders)}
    \label{fig:hierarchical-pre}
\end{figure}

The method above requires storing as many EDEKs as there are levels in the tree structure.
If we use a multi-hop, uni-directional algorithm (LWE being the only one~\cite{lwe-reencryption}), we can have re-encryption keys from the bottom levels to
the top levels~(\figref{fig:hierarchical-pre}).
When access is granted to a directory, all its files and sub-directories have a path to be re-encrypted under the directory key, and then re-encrypted for the
recipient:
\begin{equation}
    edek_b = \text{reencrypt}(re_{xb}, \text{reencrypt}(re_{1x}, edek_1)).
\end{equation}

\subsection{Encrypting bulk data}

When we have a per-file symmetric encryption key, it is possible for a malicious recipient to download all the encrypted symmetric keys and decrypt them
while he has access without the necessity of downloading overly too much data.
He can use these symmetric keys later on to decrypt the bulk of the data.
This is called a ``key scraping attack''.

In the applications where this is a problem, we can use the approach of applying all-or-nothing transformation to the data, so that one has to download all of the
file in order to decrypt it.
This approach was published~\cite{aont-bulk} and is readily available to be used in our key management system.

\subsection{Sharing encrypted streams}

Sharing encrypted streams with multiple users is ideal for applications such as a decentralized Netflix, where third parties routing traffic
aren't permitted to see the content.

The principle is simple: each block of data in the stream is encrypted with a random DEK, and EDEK is produced using a per-channel public key.
However, it takes time to make a roundtrip with \kms.
Therefore, it makes sense to also include an EDEK of the next block of data together with the previous block.

Also, for the use case of streaming media content, key scraping attack are especially problematic.
Thus, we should consider blocks of movies as bulk data and encrypt accordingly~\cite{aont-bulk}.

As usual, consumers of the stream have their own key pairs and ask miner nodes to make EDEKs decryptable by the consumers.
Miners can also ensure the consumer actually paid for the subscription and refuse to re-encrypt if they did not.

\subsection{Time-based and condition-based policies}

While we don't trust miners with encrypted data or keys, we still trust them to control the duration of storing a re-encryption key.
The simplest policy is time-based: re-encryption is allowed only during the specified time interval and, if there is no time in the future when
re-encryption is allowed, the re-encryption key should be removed.

More complex policies can be created.
Re-encryption can be allowed on a condition~--- for example, pending the completion of a certain transaction.
This allows applications like pay-per-content DRM, or storing secret data as an escrow for financial transactions.

\subsection{Key rotation}
\label{sec:key-rotation}

Many proxy re-encryption algorithms can be applied multiple times, BBS98~\cite{BBS98} included.
Thus, proxy re-encryption can be used for key rotation.
Key rotation allows all the EDEKs encrypted with an old key to become encrypted with a new key.

The owner of the data needs to produce re-encryption keys $re_{v1\rightarrow v2}$ between two versions of secret keys, essentially
sharing data with her future self.
Now, re-encryption nodes (assuming the encrypted storage is public, such as IPFS) will download EDEKs and apply the transformation.
For this operation, no collusion risks exist since the ``sender'' and ``receiver'' of the data are, essentially, the same person.

\section{Designing a smart contract}
\label{sec:smart-contract}

For the initial release, we will ensure that the nodes are staying online and correctly re-encrypting the data without requiring
anonymization~(Sec.~\ref{sec:anonymity})..
We expect the nodes to become professionalized and invest heavily in their security, similar to what's happened in other networks.
Thus, we will have a minimum stake requirement of $s_{\min}$ for nodes to provide re-encryption services.
This also naturally limits the number of nodes, similar to the mechanism for staking master nodes in the DASH cryptocurrency~\cite{dash:whitepaper}.

The amount of stake is public, so clients of the network can decide for themselves whether to deploy re-encryption keys with a node or not.
The software will automatically deploy a re-encryption key at the node which will be available by quotas, meaning:
\begin{equation}
    n_{\text{rekeys}} < (1 + \alpha) \frac{\text{stake}}{\text{total\_stake}},
\end{equation}
where $\alpha$ is a small value (for example, $\alpha=0.1$.
If the condition above is satisfied, the ``closest'' hash of the node's pubkey (the smallest Hamming distance) to the hash of the policy is chosen,
out of the remaining available nodes.
If the client decides to ignore the quotas, he risks compromising the security of his own policy (but not anyone else's), so this quota doesn't have
to be enforced by a smart contract.

In order to provide re-encryption services, a node needs to send its deposit to a smart contract (while specifying the lock time).
After the time expires, the node can withdraw the stake from it.
The objective is to have rewards minted to staking nodes if they correctly provide re-encryption services.

Let's call the suite of possible tests a generalized ``ping''.
This includes:
\begin{itemize}
    \item the actual ping (is the node still showing in the network?);
    \item check if an old but active re-encryption policy still works;
    \item check if a create-reencrypt-revoke sequence works;
    \item check if the re-encryption policies above work if requested by a different address, not related to the mining node.
\end{itemize}
When performing this ``ping'', one can choose to do only the tests required to determine availability.
For example, if a node goes offline, the (very lightweight) actual ping will be enough to determine the node's status.

Now, we want to design a system of nodes which does a self-check every $h$ blocks (where $h=10$, for example).
The idea is that nodes are betting on which nodes are misbehaving, and mine rewards depending on whether their guess is correct or not.
The ``correct'' guess is considered to be the one shown by the stake majority of the nodes, as calculated by the smart contract from votes of all the nodes.

The protocol wouldn't scale well with the total number of nodes in the system $N$ if we check every single node each $h$ blocks.
So, only a portion of the nodes (for example, $k=\sqrt{N}$ of them) will be checked.
For example, if $100,000$ nodes are online and Ethereum block time is $24$~s (current value), the health of every node will be checked approximately every $21$
hours.
The selection of the nodes to be validated can be done by the current validating nodes exchanging random bytes off-chain and calculating a hash of these
random bytes ordered and joined.
Nodes which don't participate in the validation won't be able to distinguish whether they are being challenged or receiving a real client request.

It is also not feasible for all the nodes to perform the health check.
So, a portion of nodes (for example, $\sqrt{N}$ of nodes closest to the current block hash by Hamming distance) will be able to get rewards for health checking
in the current round.
If nodes are found to be not healthy, they're punished by losing the average amount of block rewards they've obtained since the last health check.

It is possible to check a node's output for correctness of re-encryption.
In order to do that, the owner of the data can prepare a ``challenge pack'' - data which doesn't correspond to any sensitive data, but is used specifically to
challenge re-encryption nodes.
The ``challenge pack'' consists of input ciphertexts and expected re-encrypted outputs
(this method works only for proxy re-encryption algorithms which do not have probabilistic outputs).
The recipient of the data can decrypt the challenge pack and demonstrate a re-encryption node is failing or unable to properly re-encrypt data.

The nodes which challenge other re-encryption nodes are essentially voting on which nodes are misbehaving and betting on the result of this vote.
Challenger nodes should be unable to figure out the votes of others prior to everyone committing their vote.
Thus, they first commit the vote by demonstrating hashes of salted votes, later proving that they have the actual vote and salt to produce those hashes.

\section{Token economics}
\label{sec:token}

Protocol economies consist of a network of miners that contributes work to provide a scarce resource and that is
rewarded when said resource is consumed.
In \kms, miners are re-encryption nodes.
Anyone can become a miner and their rewards are differentiated based on the amount of re-encryption operations provided.
Access to the scarce re-encryption services must be controlled and allocated to the highest value uses.
The mechanism by which we achieve this is \kms~Token (NKMS).
NKMS is both the reward miners get for contributing work and the price consumers (owners of the data) pay for access to re-encryption services.
Vitally, the token also incentivizes correctness of computation and security of the system.

\subsection{Token distribution}

It is important to incentivize re-encryption operations early on, when there are not many users in the system.
That's why we'll introduce a rewards schedule where some rewards are ``mined'', asymptotically approaching zero inflation as time passes.

Miners can hold a number of re-encryption keys proportional to the number of tokens they hold as a collateral deposit, locked by a smart contract.
The miners are paid both for providing re-encryption services and making themselves available to re-encrypt data.
The miners can be paid either by the owner of the data, or by users of the data.
The latter is more relevant for usecases where DAOs distribute content under decentralized DRM.
In addition, the miners have ``availability rewards'' generated if they stay online and ready to provide their services, proof of which is checked with the help of
smart contract~(Sec.~\ref{sec:smart-contract}).

If miners become unavailable, they lose their ``availability'' compensation for the time period.

If miners cheat and provide wrong re-encryptions, they lose a fraction of their security deposit (Sec.~\ref{sec:challenge-protocol}).

We can disincentivize miners from leaking re-encryption keys.
Anyone can challenge a miner with a hash of a re-encryption key and, if the re-encryption key is proven to have been leaked, the miner forfeits his collateral
valued in NKMS.
It is possible to make re-encryption keys linked to particular proxies in order to identify them quickly if they're leaked on the encryption
level~\cite{Libert2008}.

An interesting property of \kms~is that security improves as the number of network participants grows. As additional re-encryption
nodes enter the network, the lower the chance of collusion.
This improves both the security and censorship-resistance of the system, providing powerful network effects and meaningful first mover advantages.

Finally, in order to test the security of the network, we can put bounty secrets~--- private keys for wallets with
some crypto-currency.
Anyone is free to hack the system and take the bounty.
The fact of the bounty being taken would be proof of compromised security.
This same mechanism can be used to warn of a potential data leak by individual users of the network, by signalling that they may have exposed
their key to the data.

\section{Use cases}
\kms~provides the infrastructure for a variety of applications that require sharing of sensitive data as a basic
functionality. The ability to condition decryption operations on public actions on the consensus network, such as the publication
of certain messages, payments made between specific parties, and other events, enables a range of applications including:

\subsection{Sharing encrypted files (``Decentralized Dropbox'')}
\label{sec:files}
Files can be encrypted client-side and stored in decentralized filesystems like Swarm~\cite{swarm}, IPFS~\cite{whitepaper:ipfs}, Sia~\cite{web:sia}, or Storj~\cite{web:storj}, or centralized ones like S3.
The files can be easily shared with approved third-parties by providing a re-encryption token based on the third-party's
public key.
The third-party's access permission can be easily revoked by removing the re-encryption token from the network.

\subsection{End-to-end encrypted group chat (``Encrypted Slack'')}
PRE is an ideal primitive for end-to-end encrypted group messaging, in which multiple participants require read and write
access to a channel. Members can easily be added or removed to the chat by issuing or revoking a re-encryption token.
This avoids the overhead of encrypting and sending messages multiple times, individually for each participant.

\subsection{Patient-controlled electronic health records (EHR)}
A patient-controlled EHR can be created in which the patient owns their data and encryption keys, as opposed to centralized
systems like Epic.
Again, the data can be stored centrally or in a decentralized backend.
When the patient wants to share their encrypted data with a hospital or insurance company, they issue a re-encryption token,
which grants temporary access to the third-party.

\subsection{Decentralized digital rights management (DDRM)}
\label{sec:drm}
Cryptographic access control can act as a decentralized DRM.
Access controls can be embedded into the encryption itself so that they follow the data wherever it goes.
Conditional re-encryption tokens can be controlled by a smart contract and released only upon payment.
Services like a decentralized Netflix or an encrypted marketplace selling software, apps, photos, and other digital content
can now be built using \kms.

\subsection{Blind identity management}
A blind identity management service can be constructed using \kms. Identities can be encrypted client-side and stored with the
identity management provider. Users can create re-encryption keys for approved applications. The service re-encrypts identity
credentials for said third-party applications, without the identity provider ever having access.

\subsection{Secret credentials management for scripts and backend applications}
\kms~is ideal for the storage of any secrets, such as sensitive environment variables, database credentials, and API keys.
For scripts, a re-encryption token can be generated for the duration of a script, then revoked.
For example, developers can safely store encrypted database credentials on GitHub, giving temporary access to these credentials
once an instance is deployed.
Even if the GitHub repository is public, the credentials cannot be used by an unauthorized person.

\subsection{Shared credentials and enterprise password management}
\kms~can manage shared credentials that employees use to access web services.
An audit log can be built to monitor who accesses what secrets.
When an employee leaves, it is easy to revoke access or even roll keys.

\subsection{Mandatory access logging}
In some corporate and enterprise settings, clients must publish access logs for sensitive files.
This requires that each file access be recorded and conditional re-encryption can be used to mandate these logging rules.

\subsection{Mobile device management (MDM) and revocation}
\label{sec:mdm}
In an enterprise MDM setting, re-encryption tokens can be created for valid devices. When a device is lost or retired,
or an employee leaves the organization, the re-encryption token can be deleted to revoke the device's access. This avoids the
problem of re-organizing hierarchical key trees.

\section{Summary}
\kms~is a decentralized key management service and cryptographic access control layer for the blockchain and
decentralized applications.
Developers and enterprises alike can leverage it to create highly-secure applications in healthcare,
financial services, and more.
By bringing private data sharing and computation to the public blockchain, \kms~enables everything from encrypted
content marketplaces to secret credentials management to patient-controlled electronic health records.

\section{Acknowledgements}
We would like to thank Giuseppe Ateniese from Stevens Institute of Technology and Isaac Agudo Ruiz from University of M{\'a}laga
for their help with proxy re-encryption algorithms.
We also thank Dave Evans at the University of Virginia for advising our company throughout its lifetime
and his help with regards to the current state of the art in multi-party computation.
And Stefano Bernardi, Tom Ding, and many others for their help on token economics.

\bibliography{kms-whitepaper}

\end{document}